# Thermo-reversible permanent magnets in the quasi-binary GdCo$_{5-x}$Cu$_x$ system

R.M. Grechishkin, M. S. Kustov

*Laboratory of Magnetoelectronics, Tver State University, 170000 Tver, Russia*

O. Cugat, J. Delamare, G. Poulin, D. Mavrudieva

*Laboratoire d'Electrotechnique de Grenoble, BP 46, 38402 Saint Martin d'Hères, France*

N.M. Dempsey

*Laboratoire Louis Néel, 25 avenue des Martyrs, BP 166, 38042 Grenoble, France*

Ferrimagnetic GdCo$_{5-x}$Cu$_x$ alloys exhibiting the effect of Gd- and 3$d$-sublattice magnetization compensation at defined temperatures were studied with respect to their use as thermo-reversible permanent magnets (PM). Coercive fields $\mu_0 H_c$ in the range 0.3 to 1.6 T were measured for annealed single crystals with $x = 1 - 2.2$ having compensation points in the vicinity of room temperature. Two applications of such a thermo-reversible PM, namely a thermally controlled actuator and a contactless temperature sensor, are demonstrated.

Permanent magnets (PM) are used as a source of magnetic flux in a wide range of applications.[1] In general, for most applications it is required that the magnetic properties of the magnet (remanence and coercivity) are stable over the working temperature range. On the contrary, we propose to exploit the thermal variation of the magnetic properties of ferrimagnetic alloys with a compensation temperature.[2] In this study we demonstrate that the substitution of appropriate amounts of Cu for Co in $Gd(Co_{5-x}Cu_x)$ leads to the development of significant values of coercivity over a useful temperature range, rendering it a permanent magnet.

The role of a third, 3$d$-transition, element such as Cu, Al, Ni or Fe in modifying the magnetic behavior of rare-earth cobalt alloys has been the subject of a number of studies. In particular, it was demonstrated that Cu and Ni substitutions in the quasibinary $Sm(Co_{5-x}M_x)$ system may lead to giant magnetic hardness combined with giant magnetic viscosity[3-6]. Al and Fe replacements were shown to be useful in regulating the easy plane – easy axis spin reorientation temperature region in quaternary $Nd(Co_{1-x-y}Al_xFe_y)_5$ soft magnetic alloys applicable in thermal sensors and actuators,[7] while Ni substitutions for Co provided an effective control of the compensation temperature in ferrimagnetic Gd-Co-Ni films.[4]

In the present study we investigated the ferrimagnetic quasibinary $Gd(Co_{5-x}Cu_x)$ system in the expectation that Cu substitution will, by analogy with $Sm(Co_{5-x}Cu_x)$ compounds, increase the coercive force of the magnetically soft binary GdCo$_5$ to values typical of PM (i.e. $H_c \geq 2\pi M_s$), retaining at the same time the phenomena of reversible macroscopic compensation of the material's magnetization at some predetermined temperature.

Coarse-grained (average grain size 2 – 4 mm) 50 g ingots were prepared by induction melting in alumina crucibles of Gd, Co (both of 99.8% purity) and Cu (99.99%). The ingots were vacuum annealed at 1273 K for 2 hours and then cooled to room temperature (RT) at a rate of 100 K/min. No attempts were made for special heat treatments which possibly will be useful in maximizing the coercive field. Spherical single crystal samples (diameter ~2 mm) were ground from selected large grains. Hysteresis loops were measured using an automated VSM in an electromagnet with a maximum field of 3 T. Uniaxial Magneto-optic indicator films (MOIF)[8] were used to visualize the stray fields and magnetic domain structures of the samples.

Shown in figure 1 are the RT values of the saturation magnetization and coercive field for $GdCo_{5-x}Cu_x$ ($x = 0 - 2.5$) samples as a function of composition. It is seen that the RT value of the compensation point and the maximum measured value of $H_c$ occur at $x \approx 1.5$; coercive fields for the practically interesting temperature interval in the vicinity of RT are much greater than in the binary $GdCo_5$ ($x = 0$, $\mu_0 H_c \approx 30$ mT), varying between $0.3 - 1.6$ T within the range $x = 1 - 2.2$.

Figure 2 presents the room temperature major hysteresis loops of electropolished[9] single crystals for the cobalt-rich part of the $GdCo_{5-x}Cu_x$ system ($x = 0.25 - 1.25$) measured after saturation in fields of $\pm 2$ T (shown in the graph are the central parts limited to $\pm 0.8$ T). These loops demonstrate 100% remanence and ideal rectangularity, similar to those observed earlier in binary alloys[10]. In non-polished polycrystalline samples the remanence and loop rectangularity depend on the degree of texture; the coercive field values, however, are the same or higher than in single crystals. The coercive field $H_c$ increases and the saturation magnetization $\sigma_s$ decreases as $x$ approaches the compensation composition ($x = 1.5$). The behavior is reversed as $x$ increases above this value (hysteresis loops not shown for clarity).

Direct evidence of magnetization inversion on passing through the compensation point was obtained by visualization of the sample's stray field with the aid of a uniaxial MOIF. The (a) and (b) images of figure 3 show the distribution of the normal component of the field at the surface of the $GdCo_3Cu_2$ sample magnetized to saturation. The sample, in the shape of a thin disk with a diameter of 2 mm and thickness 0.1 mm, was cut from an oriented spherical crystal in such a way that the easy axis of magnetization (crystallographic $c$-axis) was normal to the surface. It is seen that at RT the sample is represented by a bright uniform circle in the center of the MOIF image surrounded by a dark ring representing, as expected, opposite directions of the stray field directly above and beside the sample, respectively. The whole picture is inverted at 100 $^{o}$C, reflecting the change in dominance from the Gd sublattice magnetization to that of the 3$d$ sublattice, on passing through the compensation point $T_{comp}$ = 90 $^{o}$C. Heating and cooling cycles were repeated many times without any signs of magnetization degradation. During temperature cycling, the MOIF shows its own unaffected domain structure at $T_{comp}$, demonstrating the zero magnetization state of the sample at this point.

The (c) and (d) images presented in figure 3 illustrate the effect of the temperature on the same sample in its multidomain state obtained by DC demagnetization. In these cases a 180$^{o}$ magnetic domain structure typical of uniaxial crystals is clearly observed at temperatures below and above $T_{comp}$. Passing through $T_{comp}$ results in an inversion of the domain contrast. In the vicinity of $T_{comp}$ the magnetization of the sample is near to zero, so the uniaxial MOIF demonstrates its own intrinsic stripe domain structure which is practically unaffected by the sample under study (Fig. 3(e)). It is noteworthy that the configuration of the inverted domain structure reproduces the non-inverted one with a high degree of accuracy. This behavior suggests a unique possibility of producing a multipole magnet defined by artificially engineered periodic or aperiodic frozen domain structure configurations. In addition to the possibility of thermally switching between north and south pole configurations, such a magnet will also possess a very narrow (nanosize) transition zone width, equal to the domain wall width, between opposite poles. This feature is potentially important for applications in micro- and nano-electro-mechanical-systems (MEMS and NEMS). Furthermore, the possibility of changing the magnetization pattern (the position of transition zones) through the application of non-uniform temperature fields may also be exploited.

On the macroscopic scale, the effect of magnetization reversal with temperature was demonstrated by heating the sample suspended by a thread in the field of a permanent magnet. On passing through the compensation point, the sample rotates by 180$^{o}$ and oscillates for some time in a decaying fashion, with a frequency depending on the moment of inertia of the system and rigidity due to both the sample and the suspension. This experiment serves to demonstrate two potential applications for such materials: the thermally induced sample reversal may be exploited in an actuator while the oscillations could be exploited in an original resonant temperature sensor.

As another demonstration of the potential application of such a thermoreversible PM, we developed a modification of the magnetic temperature sensors described recently by Fletcher and Gerschenfeld.[11] These sensors are comprised of three basic elements: the *signal element* (square loop soft magnetic tape producing detectable harmonics or detectable resonance frequency), the *bias element* (thin strip of semi-PM material), and the *modulation element* (soft magnetic material with a Curie temperature near the operating temperature for a given application). The modulation element changes its magnetic permeability with temperature leading, in turn, to a change of the biasing field of the adjacent semi-PM material acting on the signal element. As a result the magnetization harmonic spectrum of the signal element varies with temperature. This variation can be detected over a distance of several inches.

The same principle of temperature measurement may be implemented making use of one thermoreversible PM instead of two (bias and modulation) elements in the original Fletcher and Gerschenfeld sensor. When the signal element (nanocrystalline $Fe_{81}B_{13.5}Si_{3.5}C_2$ alloy in the form of a $40\times5\times0.02$ thin strip in our case) is interrogated by an AC external field, the output signal spectrum in the absence of DC bias is presented by odd harmonics only (figure 4(a)). DC biasing by a small (100 mg) piece of $GdCo_3Cu_2$ PM placed at a distance of several millimeters from the strip results in the appearance of even harmonics with an amplitude comparable to that of the odd ones (figure 4(b)). The temperature dependence of the second harmonic amplitude presented in the inset of figure 4b is rather linear and completely reversible. It may be noted that the slope of the $M_r(T)$ curve of such a thermo-reversible PM is not changed when approaching the compensation point, thus allowing clear zero reading, while ordinary thermomagnetic materials may suffer from magnetization "tails" near the Curie point.

To summarize, the present work has demonstrated that it is possible to have a combination of both ferrimagnetic compensation type behavior and magnetic hardness in $GdCo_{5-x}Cu_x$ alloys. Despite the relatively low values of magnetization, these alloys may be used as a special kind of PM, the magnetization of which can be reduced to zero or reversed by varying temperature. Experiments have been carried out to demonstrate the potential application of such a material in contactless temperature sensors as well as thermally controlled actuators. Other applications in magnetic MEMS are envisaged. Due to the resemblance of magnetic and physical properties of rare earth – transition metal alloys, other compositions with similar or better properties are likely to be found and alternative preparation methods such as powder metallurgy[7] or thin-film technologies may be applicable.

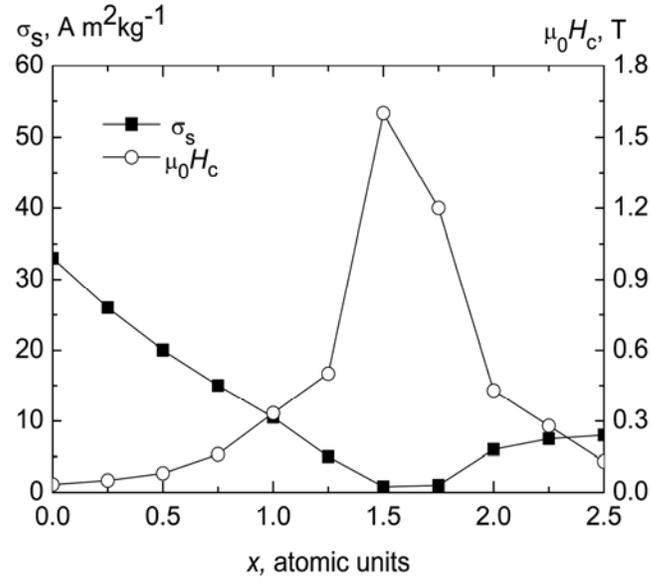

Fig. 1. Room temperature specific saturation magnetization $\sigma_s$ and coercive field $\mu_0H_c$ of GdCo$_{5-x}$Cu$_x$ ($x = 0 - 2.5$) single crystals as a function of composition.

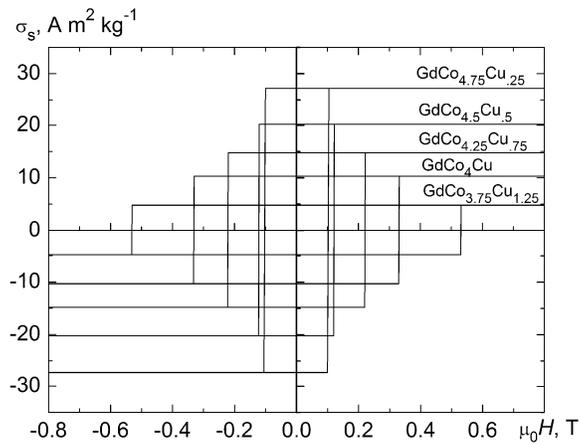

Fig.. 2. Room temperature easy axis major hysteresis loops of electropolished spherical single crystals of GdCo$_{5-x}$Cu$_x$ ($x = 0.25 - 1.25$).

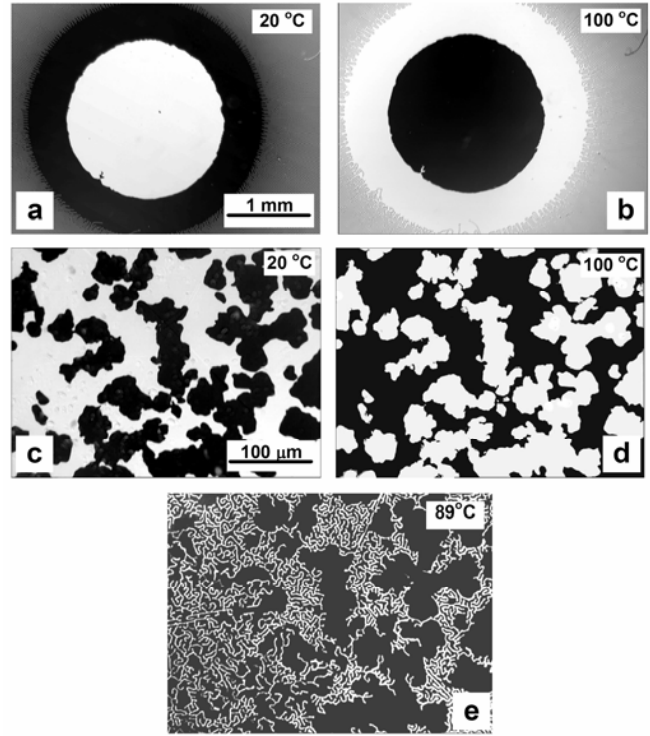

Fig. 3. Temperature-induced inversion of magnetization of saturated ((a), (b)) and demagnetized ((c), (d)) 2 mm disk GdCo$_3$Cu$_2$ single crystal as observed with a uniaxial MOIF at temperatures below and above the compensation temperature $T_{comp}$ = 90 $^o$C. The intrinsic stripe domain structure of the uniaxial MOIF is seen in the vicinity of $T_{comp}$ (e).

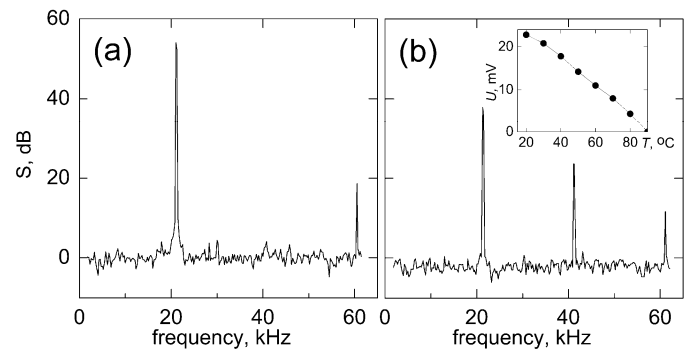

Fig. 4. Room temperature output signal spectrum of the sensor. (a) AC sinusoidal excitation at a frequency of 20040 Hz with zero bias (1$^{st}$ and 3$^{rd}$ harmonics); (b) same as (a) after DC biasing of the signal element by a thermo-reversible GdCo$_3$Cu$_2$ permanent magnet (2$^{nd}$ harmonic appeared). Inset: temperature dependence of the 2$^{nd}$ harmonic amplitude.